\documentstyle[axodraw,iopconf1,epsf]{article}
\begin{document}

\begin{flushright}
MPI/PhT/98-75\\
hep-ph/9810211\\
October 1998
\end{flushright}

\title{Baryogenesis through mixing of heavy Majorana neutrinos}

\author{Apostolos Pilaftsis\footnotetext[1]{To appear in the   proceedings
  of 2nd International Conference on Dark Matter in Astro and Particle
  Physics  (DARK98),   ed.\ H.V.  Klapdor-Kleingrothaus,   Heidelberg,
  Germany, 20-25 July 1998.}}

\affil{Max-Planck-Institut f\"ur Physik, F\"ohringer Ring 6,\\
 80805 Munich, Germany, and\\
Theory Division, CERN, CH-1211 Geneva 23, Switzerland}

\beginabstract  
We review the scenario of baryogenesis through leptogenesis induced by
the out-of-equilibrium   decays  of heavy  neutrinos.   We pay special
attention to the resonant phenomenon of CP violation through mixing of
two nearly degenerate heavy Majorana neutrinos  and show how unitarity
and CPT  invariance is maintained  within the resummation approach.  An
important consequence of this is that the leptogenesis scale may be as
low as  1 TeV, even  for models with universal   Yukawa couplings.  We
briefly discuss   the    impact of   finite  temperature  effects  and
low-energy  constraints   to  the  afore-mentioned    mechanism of  CP
violation.  
\endabstract

\section{Introduction}

Many   scenarios  have been  invoked  in   order to  explain the small
baryon-to-photon ratio of number densities $n_B/n_\gamma = (4-7)\times
10^{-10}$,  which is found by   present observations.  Among them, the
most attractive scenario is the  one proposed by Fukugita and Yanagida
\cite{FY}  which may lead to a  consistent  solution to the problem of
the baryon asymmetry  in     the Universe  (BAU).  In  their    model,
out-of-equilibrium $L$-violating   decays of heavy  Majorana neutrinos
$N_i$ with  masses  $m_{Ni}$  much  larger  the  critical  temperature
$T_c\approx  200$  GeV produce an  excess in  the lepton   number $L$. 
Then, the excess in $L$  is converted into  the desired $B$  asymmetry
through $B+L$-violating sphaleron  interactions \cite{KRS}, which  are
in thermal  equilibrium  for  temperatures ranging  from   $T_c$ up to
$10^{12}$  GeV \cite{AMcL,BS}. Over the last  years, many studies have
been devoted to   this mechanism of  baryogenesis through leptogenesis
\cite{MAL,CEV,epsilonprime,LS,Paschos,APRD}.

In \cite{FY},  the necessary CP  violation in  heavy Majorana neutrino
decays results from the interference between  the tree-level graph and
the  absorptive   part  of the   one-loop vertex.    Using   the known
terminology of the $K^0\bar{K}^0$ system \cite{reviewCP}, we call this
mechanism of CP violation   as $\varepsilon'$-type CP violation.   The
$\varepsilon'$-type CP  violation  was  discussed extensively  in  the
literature \cite{MAL,CEV,epsilonprime}.  Provided all Yukawa couplings
of the   Higgs fields  to $N_i$  and  lepton   isodoublets $L$  are of
comparable  order  \cite{MAL,epsilonprime}, baryogenesis  through  the
$\varepsilon'$-type mechanism requires  very heavy  Majorana neutrinos
of  order $10^7-10^8$ GeV.  This  high mass bound  may be  lifted if a
strong hierarchy for  the Yukawa   couplings  and the  heavy  Majorana
neutrino  masses  is   assumed   \cite{MAL,CEV}.  Without  the   above
assumption  on  the Yukawa couplings however  \cite{epsilonprime}, one
finds that $\varepsilon'<10^{-15}$ for $m_{Ni}\approx 1$ TeV, which is
far too small to account for the BAU.

In heavy    particle  decays responsible   for the    baryon  (lepton)
asymmetry, CP  violation may also arise  from the interference  of the
tree-level graph with the absorptive part  of a self-energy transition
describing      the       mixing     of     the  decaying     particle
\cite{IKS,KRS,LS,Paschos,APRD}.  We call the latter $\varepsilon$-type
CP violation.  Recently, it has  been observed in  \cite{Paschos,APRD}
that  CP violation can be considerably  enhanced through the mixing of
two nearly degenerate heavy Majorana  neutrinos.  A minimal model with
nearly degenerate neutrinos may be  the low-energy  limit of E$_6$  or
SO(10) theories. We will discuss such a scenario in Section 2.
 
In order  to describe the dynamics  of  CP violation  through unstable
particle  mixing, one is  compelled to rely on resummation approaches,
which treat  unstable particles in  a consistent way and  preserve all
desirable   field-theoretic  properties   such    as gauge-invariance,
analyticity, CPT invariance and  unitarity.  In  the context of  gauge
field theories,  such  a resummation approach   to resonant transition
amplitudes has been  formulated,  which is  implemented by the   pinch
technique \cite{PP}.  Then, this formalism   has been extended to  the
case of mixing between two intermediate  resonant states in scattering
processes  \cite{APRL,AP}.  In \cite{APRD},   a related formalism  for
decays  has been developed, which  effectively takes into account {\em
  decoherence}  phenomena in the mixing  and subsequent decay of heavy
particles,  namely heavy  neutrinos.   Within this formalism, we  have
shown that $\varepsilon$-type CP violation can be  even of order unity
\cite{APRD}. Thus, we find agreement with earlier articles on resonant
CP violation  in  scatterings involving   top quarks,   supersymmetric
quarks     or  Higgs    particles    in    the   intermediate    state
\cite{APCP,APRL,AP}.

In Section 3, we will briefly describe the afore-mentioned resummation
approach and show   how unitarity and  CPT invariance   is maintained. 
Based on this approach, we  will calculate the leptonic CP asymmetries
of heavy neutrinos and derive  the sufficient and necessary conditions
for order-unity  CP  violation. As we  will see,  this very last  fact
allows one  to  lower  the scale   of leptogenesis  up  to TeV  scales
\cite{APRD},  even   for models   without  any  large  Yukawa-coupling
hierarchy.  In  Section 4, we discuss  the implications  of low-energy
constraints such as the electric  dipole moment (EDM) of electron  for
the  leptogenesis  scenario.  In Section  5,  we examine the impact of
finite-temperature effects on the resonant  mechanism of CP  violation
and  give predictions of the baryonic  asymmetry in two representative
scenarios, after solving numerically the relevant Boltzmann equations.
Section 6 contains the conclusions.

\section{Heavy Majorana neutrino models}

Heavy Majorana  neutrinos may naturally be  realized in certain GUT's,
such as SO(10) \cite{FM,Wol/Wyl} and/or  E${}_6$ \cite{witten} models. 
Nevertheless, these models  will also predict several other particles,
{\em e.g.}, leptoquarks, additional  charged and neutral  gauge bosons
($W^\pm_R$ and  $Z_R$), which may deplete  the number density of heavy
neutrinos  $N_i$ through  processes   of the  type  $N_i\bar{e}_R  \to
W^{+*}\to   u_R   \bar{d}_R$ and so  render   the  whole analysis very
involved.   If these  particles  are   sufficiently heavier  than  the
lightest   heavy  Majorana neutrino   and/or   the temperature  of the
universe \cite{MAL}, this problem  may be completely avoided. Since we
wish to simplify our analysis without sacrificing any of the essential
features involved in the study of the BAU, we shall consider a minimal
model with isosinglet neutrinos, which is invariant under the SM gauge
group, SU$(2)_L\otimes$U$(1)_Y$. To this end,   we consider a  minimal
scenario,  in  which the fermionic matter   of the SM  is  extended by
adding two right-handed neutrinos  per family, {\em  e.g.}, $\nu_{lR}$
and $(S_{lL})^C$  with $l=e,\mu,\tau$. Such  a scenario may be derived
from certain SO(10)  \cite{Wol/Wyl} and/or E$_6$ \cite{witten} models. 
For  our  illustrations, we  neglect  possible  inter-family  mixings. 
Imposing  lepton-number conservation  on the  model  gives rise to the
Lagrangian
\begin{equation}
\label{WW}
- {\cal L}\ =\ \frac{1}{2}\, (\bar{S}_L,\ (\bar{\nu}_R)^C )\, 
\left( \begin{array}{cc}
0 & M \\
M & 0 \end{array} \right)\, \left( \begin{array}{c}
(S_L)^C \\ \nu_R \end{array} \right)\ +\ 
h_R\, (\bar{\nu}_L,\ \bar{l}_L) \tilde{\Phi} \nu_R\ +\ \mbox{H.c.},
\end{equation}
where  $\tilde{\Phi}=i\sigma_2\Phi$  is  the isospin  conjugate  Higgs
doublet and $\sigma_2$ is the  usual Pauli matrix.   There are now two
equivalent ways to break both $L$ and CP  invariance of the Lagrangian
in   Eq.\ (\ref{WW}):  One has  to  either  (i) introduce  two complex
$L$-violating mass  terms  of the kind  $\mu_R\bar{\nu}_R\nu^C_R$  and
$\mu_L\bar{S}^C_L S_L$, where both $\mu_R$ and $\mu_L$ are complex, or
equivalently (ii) add the $L$-violating coupling $h_L\, (\bar{\nu}_L,\ 
\bar{l}_L)  \tilde{\Phi} (S_L)^C$  and  include one $L$-violating mass
parameter, {\em e.g.}, $\mu_R\bar{\nu}_R\nu^C_R$.  The origin of these
parameters are usually due   to residual effects   of high-dimensional
operators involving  super-heavy neutrinos \cite{witten}.  The typical
size  of  the   $L$-violating  parameters   is  $\mu_L,\ \mu_R    \sim
M^2/M_{Planck}$,  $M^2/M_X$ or $M^2/M_S$,  where $M_S\approx 10^{-3}\,
M_X$ is   some  intermediate see-saw scale.    As a  consequence, such
effective minimal models  derived  from E$_6$ theories  can  naturally
predict small mass splittings for the heavy neutrinos $N_1$ and $N_2$.
To a good approximation, this small  mass difference may be determined
from the parameter $x_N=m_{N_2}/m_{N_1}-1\sim  \mu_L/M$ or $\mu_R/M$.  
For  instance,  if $M=10$ TeV  and $\mu_L=\mu_R  =  M^2/M_X$, one then
finds $x_N\approx 10^{-12} -  10^{-11}$. As we will  see in Section 4,
these small values for the mass difference  $x_N$ can produce large CP
asymmetries in the heavy neutrino decays.

Using generalized  CP  transformations \cite{BBG}, one  can derive the
sufficient and necessary condition for CP invariance in a general weak
basis of the model. The condition is given by \cite{APRD}
\begin{eqnarray}
\label{CPinv}
\Im m\, \mbox{Tr} ( h^\dagger h M^{\nu\dagger} M^\nu M^{\nu\dagger}
h^T h^* M^\nu ) && \nonumber\\
&& \hspace{-3cm}=\ m_{N_1}m_{N_2} (m^2_{N_1} - m^2_{N_2})\,
\Im m (h_{l1}h^*_{l2})^2\ =\ 0\, .
\end{eqnarray}
where $h = (h_{l1}, h_{l2})$ is a $1\times 2$ Yukawa matrix defined in
the  mass   basis, {\em  i.e.},  $M^\nu=\widehat{M}^\nu$.   From  Eq.\ 
(\ref{CPinv}), one  readily sees  that only one  physical CP-violating
combination is  possible in this minimal  model  and CP  invariance is
restored   if   $m_{N_1}=m_{N_2}$  provided   none  of the  isosinglet
neutrinos  is massless.  The above  considerations may  be extended to
models with more than  two  right-handed neutrinos  and more than  one
lepton families. In this case,  there may be more conditions analogous
to Eq.\   (\ref{CPinv}),  which  involve  high  order   terms  in  the
Yukawa-coupling  matrix $h$.  However,  not  all of the conditions are
sufficient and  necessary  for CP invariance.  Instead  of undertaking
the rather difficult task   to derive all CP-invariant conditions,  we
note in  passing  that the    total number  ${\cal  N}_{CP}$  of   all
non-trivial CP-violating phases in a model with $n_L$ weak isodoublets
and  $n_R$  neutral  isosinglets  is   ${\cal  N}_{CP} =   n_L(n_R-1)$
\cite{KPS}.

\section{Resummation approach: CP asymmetries and CPT invariance}

Here, our main  concern is to  present an approach to decay amplitudes
that describes the dynamics  of unstable particle mixing.   Because of
the many decoherent collisions of  the heavy particle with the thermal
bath in the early universe \cite{LES},  we must take into account only
the incoherent part of the mixing and decay  of the unstable particle.
This   will be   done in   an effective  manner,   such  that the {\em
  incoherent} decay amplitude derived with this method can be embedded
in an  equivalent  form   to a  transition   element  \cite{PP,AP}  in
agreement with Veltman's S-matrix approach \cite{Velt}.

Our  approach is inspired by  the LSZ reduction  formalism \cite{LSZ}. 
Suppose that we wish  to find the  effective resummed amplitude ${\cal
  T}_{N_1}$ for the decay  $N_1\to   L \Phi^\dagger$ which    includes
$N_1N_2$ mixing. Then, we start with  the Green function $\Phi^\dagger
L  N_1$  shown in  Fig.\  1 and  amputate the  external  legs by their
inverse propagators. For the stable-particle lines $L$ and $\Phi$, the
procedure does not differ from the usual  one known from field theory. 
For the external line describing the $N_1N_2$ system, one has first to
consider the resummed propagator of the $N_1N_2$ system
\begin{equation}
\label{InvS}
S_{ij}(\not\! p)\ =\ \left[ \begin{array}{cc} \not\! p - m_{N_1} +
\Sigma_{11}(\not\! p) & \Sigma_{12}(\not\! p)\\
\Sigma_{21}(\not\! p) & \not\! p - m_{N_2} + \Sigma_{22}(\not\! p)
\end{array} \right]^{-1},\qquad
\end{equation}
where   $m_{N_{1,2}}$  are the    heavy neutrino  masses, $\Sigma_{ij}
(\not\!\!  p)$ are the $N_j\to  N_i$ self-energy transitions which are
renormalized in the  on-shell  (OS) renormalization scheme \cite{KP}.  
Then, as we did for the stable-particle lines, we have to truncate the
resummed $N_iN_j$  propagator   in Eq.\  (\ref{InvS})  by the  inverse
$N_1N_1$ propagator. To this end, we need to calculate the propagators
\begin{eqnarray}
\label{S11}
S_{11}(\not\! p) \!\!\!&=&\!\!\! \Big[\not\! p\, -\, m_{N_1}\, +\,
\Sigma_{11}(\not\! p)\, -\, \Sigma_{12}(\not\! p)
\frac{1}{\not\! p - m_{N_2} + \Sigma_{22}(\not\! p)}
\Sigma_{21}(\not\! p) \Big]^{-1}\quad \\
\label{S21}
S_{21}(\not\! p) \!\!\!&=&\!\!\! -\, \Big[
\not\! p\, -\, m_{N_2}\, +\, \Sigma_{22}(\not\! p)
\Big]^{-1} \Sigma_{21}(\not\! p)\, S_{11}(\not\! p)\ .
\end{eqnarray}
\begin{center}
\begin{picture}(300,100)(0,0)
\SetWidth{0.8}

\Vertex(50,50){2}
\Line(50,50)(90,50)\Text(70,62)[]{$N_1$}
\Line(130,50)(170,50)\Text(150,62)[]{$N_i$}
\GCirc(110,50){20}{0.9}\Text(110,50)[]{{\boldmath $\varepsilon$}}
\GCirc(180,50){10}{0.9}\Text(180,50)[]{{\boldmath $\varepsilon'$}}
\DashArrowLine(187,55)(220,80){5}\Text(225,80)[l]{$\Phi^\dagger$}
\ArrowLine(187,45)(220,20)\Text(225,20)[l]{$L$}

\end{picture}\\
{\small {\bf Fig.\ 1:} ~$\varepsilon$- and  $\varepsilon'$-type CP violation
in the decays of heavy Majorana neutrinos.}
\end{center}

Employing now the LSZ reduction procedure, we obtain the effective
decay amplitude
\begin{eqnarray}
\label{TN1}
{\cal T}_{N_1} \!\!&=&\!\! 
{\cal T}^{amp}_{N_i}\, S_{i1}(\not\! p)\, S^{-1}_{11}(\not\! p)\, 
u_{N_1}(p)\nonumber\\
\!\!&=&\!\! \bar{u}_l P_R\, \Big\{ 
h_{l1}+i{\cal V}^{abs}_{l1} (\not\! p) -
i\Big[h_{l2}+i{\cal V}^{abs}_{l2} (\not\! p) \Big]\nonumber\\  
&&\times \Big[\not\! p - m_{N_2} + i\Sigma_{22}^{abs}(\not\! p)\Big]^{-1} 
\Sigma_{21}^{abs}(\not\! p)\Big\} u_{N_1} ,
\end{eqnarray}
where ${\cal T}^{amp}_{N_i}$   is the OS renormalized $\Phi^\dagger  l
N_i$   vertex,      $\Sigma_{ij}^{abs}(\not\!\!    p)$    and   ${\cal
  V}^{abs}_{li}(\not\!\!   p)$  denote the   absorptive  parts of  the
one-loop   self-energy  transition   $N_j\to  N_i$   and  the one-loop
irreducible  vertex $\Phi^\dagger lN_i$, respectively. The self-energy
and vertex functions are given by
\begin{eqnarray}
  \label{Sigabs}
\Sigma^{abs}_{ij} (\not\! p) &=& \frac{h_{li}h^*_{lj}}{16\pi}\
\not\! p P_L\ +\ \frac{h^*_{li}h_{lj}}{16\pi}\ \not\! p P_R\, ,\nonumber\\
  \label{eps'lN}
{\cal V}^{abs}_{li}(\not\! p) &=& -\, \frac{h^*_{l'i}h_{l'j}h_{lj}}{
16\pi\sqrt{p^2}}\ \not\! p P_L\, f\Big(\frac{m^2_{Nj}}{p^2}\Big)\, ,
\end{eqnarray}
with       $P_{R(L)}    =   [1   +       (-)   \gamma_5]   /    2$ and
$f(x)=\sqrt{x}[1-(1+x)\ln(1+1/x)]$ \cite{FY}.  The  CP-conjugate decay
amplitude $\overline{{\cal  T}}_{N_1}$  for $N_1\to  L^C\Phi$  may  be
obtained by taking the complex conjugate  for the Yukawa couplings and
replacing $P_R$ with $P_L$  in Eq.\ (\ref{TN1}).  Analogously, one can
find ${\cal T}_{N_2}$ and $\overline{{\cal T}}_{N_2}$.

Our main interest is now in the physical CP-violating observables
\begin{equation}
  \label{deltaNi}
\delta_{N_i}\ =\ \frac{\Gamma (N_i\to L\Phi^\dagger )\, -\, 
\Gamma (N_i\to L^C \Phi)}{\Gamma (N_i\to L\Phi^\dagger )\, +\, 
\Gamma (N_i\to L^C \Phi)}\ , \qquad \mbox{for}\ i=1,2\, .
\end{equation}
Since we wish to study the dominant  mechanisms of CP violation in the
heavy  neutrino   decays, we can    define the CP-violating quantities
$\varepsilon_{N_i}$   and    $\varepsilon'_{N_i}$  analogous   to Eq.\ 
(\ref{deltaNi}) by considering only the self-energy or only the vertex
contribution to $\delta_{N_i}$. Then,  for the vertex  contribution to
CP asymmetries, we find the known result \cite{FY}
\begin{equation}
\label{eps'}
\varepsilon'_{N_1}\ =\ \frac{\Im m(h^*_{l1}h_{l2})^2}{
8\pi |h_{l1}|^2 }\ f\Big(\frac{m^2_{N_2}}{m^2_{N_1}}\Big)\, .
\end{equation}
In the  degenerate  heavy-neutrino mass limit   [$m_{N_1} =  m_{N_2} =
m_N$],    $\varepsilon'_{N_1}$   contribution      cancels    against
$\varepsilon'_{N_2}$,   so  no net CP  violation  can  be generated in
agreement with the condition in Eq.\ (\ref{CPinv}). 
\begin{figure}[hb]
   \leavevmode
 \begin{center}
   \epsfxsize=10.cm
   \epsffile[0 0 539 652]{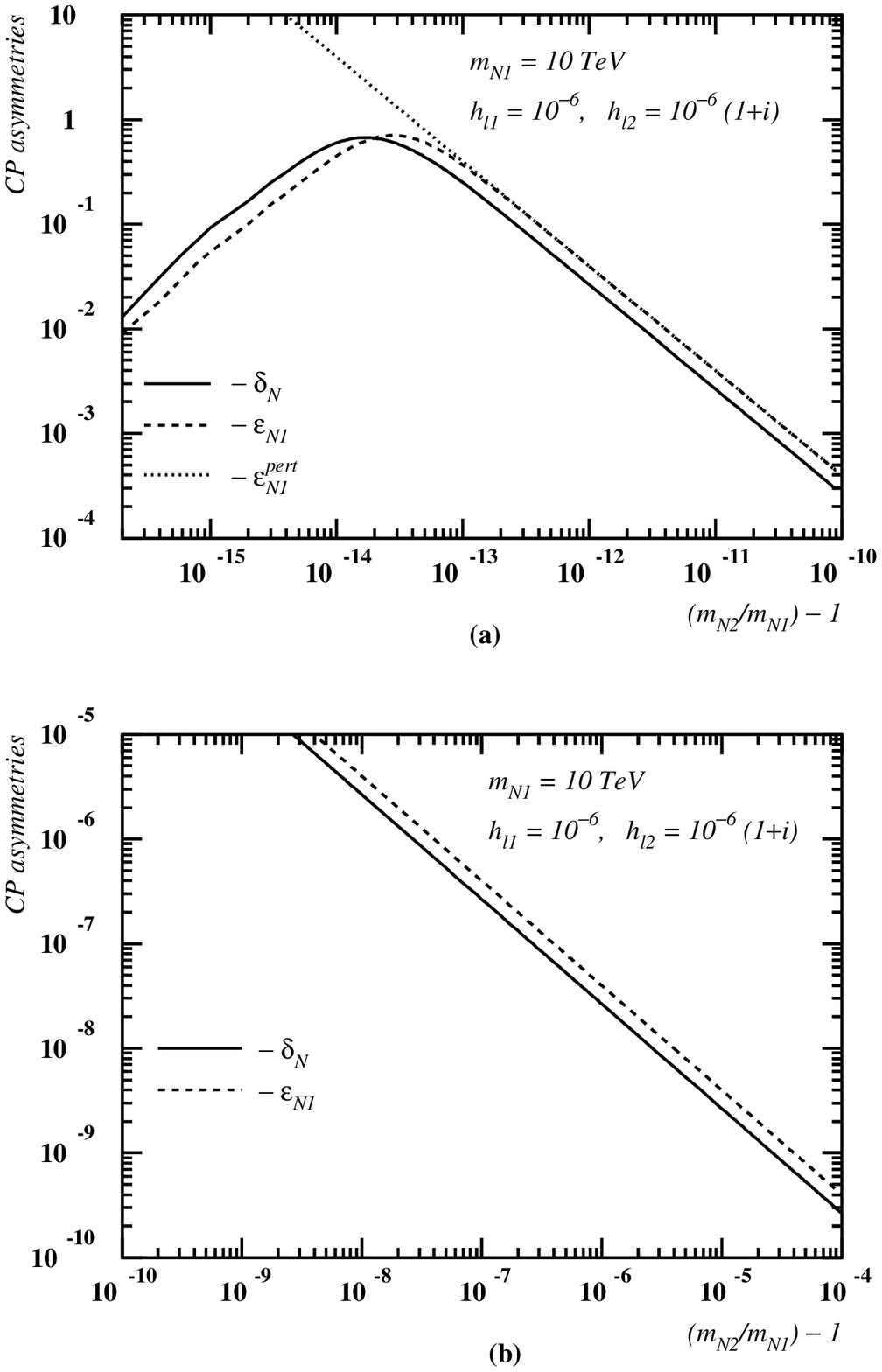}
{\small {\bf Fig.\ 2:} Dependence of CP asymmetries on $\Delta m^2_N$.}
 \end{center}
\end{figure}

Based on  the  resummation approach outlined  above ({\em  c.f.}\ Eq.\ 
(\ref{TN1})), we can calculate the  contribution which is entirely due
to   heavy-neutrino self-energy effects.   In   this way, we find  the
simple formula \cite{APRD}
\begin{equation}
\label{epsN1}
\varepsilon_{N_1}\ \approx\ \frac{\Im m ( h^*_{l1}h_{l2})^2}{
|h_{l1}|^2|h_{l2}|^2}\ 
\frac{\Delta m^2_N m_{N_1} \Gamma_{N_2} }{(\Delta m^2_N)^2\, +\, 
m^2_{N_1}\Gamma^2_{N_2}}\, ,\\
\end{equation}
where  $\Delta m^2_N  = m^2_{N_1} -   m^2_{N_2}$ and $\Gamma_{N_i}\ =\ 
|h_{li}|^2 m_{N_i}/(8\pi)$ are the decay  widths of the heavy Majorana 
neutrinos.  Equation (\ref{epsN1}) is  a  very good approximation  for
any  range of   heavy-neutrino   masses  of interest.    An  analogous
expression can  be  obtained for $\varepsilon_{N_2}$  just by changing
the  indices from  $1$ to $2$  and  vice versa in Eq.\  (\ref{epsN1}). 
Both CP asymmetries $\varepsilon_{N_1}$ and $\varepsilon_{N_2}$ are of
the same sign and go individually to zero when $\Delta m^2_N\to 0$, as
it  should be on  account of Eq.\  (\ref{CPinv}).  In the conventional
perturbation  theory, the   width  term   $m^2_{N_1}   \Gamma^2_{N_2}$
occurring in the last denominator on the RHS  of Eq.\ (\ref{epsN1}) is
absent,  and this very  last fact is precisely  what causes a singular
behaviour when the degeneracy between the two heavy Majorana neutrinos
is exact.  On physical  grounds,   however, this  should not  be  very
surprising, since the only natural parameter  that can regulate such a
singularity  is the  finite   width of  the  heavy  neutrinos which is
naturally implemented within the resummation approach.

{}From Eq.\ (\ref{epsN1}), we  can derive the sufficient and necessary
conditions   for maximal CP  violation.    Thus,  we have a   resonant
enhancement of CP violation of order unity {\em iff}
\begin{equation}
\label{CPcond}
m_{N_1}\, -\, m_{N_2}\ \sim\ \pm\, A_{22} m_{N_2}\, 
=\, \frac{\Gamma_{N_2}}{2}\
\ \mbox{and/or}\quad A_{11} m_{N_1}\, =\, \frac{\Gamma_{N_1}}{2}\ . 
\end{equation}
{\em and}
\begin{equation}
\label{dCP}
\delta_{CP}\ =\ \frac{|\Im m (h^*_{l1}h_{l2})^2|}{|h_{l1}|^2
  |h_{l2}|^2}\ \approx 1\ .
\end{equation}
A numerical example of resonant CP violation  is displayed in Fig.\ 2,
where $\varepsilon^{pert}_{N_1}$ is  the  result in   the conventional
perturbation theory and  $\delta_N$ comprises  the total  CP-violating
effect of both $N_1$ and $N_2$ decays.

It is now interesting to demonstrate  how CPT invariance and unitarity
is preserved within  the  resummation approach \cite{AP}.  As  we will
see, an immediate consequence of unitarity and CPT symmetry is that CP
violation in the  $L$-violating scattering process  $L\Phi^\dagger \to
L^C\Phi$  is  absent  to  order   $h^6_{li}$  \cite{RCV,BP}.  We  will
concentrate  on  the resonant  part of   the   amplitude which  is the
dominant one. Thus, we wish to show that to `one loop',
\begin{equation}
  \label{DCP}
\Delta_{\rm CP}\ =\ \int d{\rm LIPS}\ 
|{\cal T}^{\rm res}_{L\Phi^\dagger \to  L^C\Phi}|^2\
-\ \int d{\rm LIPS}\
|\overline{\cal T}^{\rm res}_{L^C\Phi \to  L\Phi^\dagger}|^2\ =\ 0
\end{equation}
where LIPS  stands  for the two-body Lorentz   invariant phase space.  
Furthermore,  omitting external    spinors and   absorbing  irrelevant
constants in the   definition  of the   Yukawa-coupling matrix  $h   =
(h_{l1},h_{l2})$, we have in matrix notation
\begin{equation}
  \label{Tres}
{\cal T}^{\rm res}_{L\Phi^\dagger \to  L^C\Phi}\ =\ 
hP_R\, S(\not\! p)\, P_R h^T\, ,\qquad
\overline{\cal T}^{\rm res}_{L^C\Phi \to  L\Phi^\dagger}\ =\ 
h^*P_L\, \bar{S}(\not\! p)\,P_L h^\dagger\, ,
\end{equation}
with  $\bar{S}(\not\!    p)  =  S^T(\not\!   p)$  being  the
CP/T-conjugate propagator matrix  of $S (\not\!  p)$.\\  We  start our
proof by noticing first that as a consequence of CPT invariance,
\begin{equation}
  \label{CPT}
|{\cal T}^{\rm res}_{L\Phi^\dagger \to  L\Phi^\dagger}|^2\ =\ 
|{\cal T}^{\rm res}_{L^C\Phi \to  L^C \Phi}|^2\, ,
\end{equation}
since
\begin{equation}
|hP_R\, S(\not\! p)\, P_Lh^\dagger|\ =\ |h^*P_L\, S^T(\not\! p)\, P_R h^T|\
=\ |h^*P_L\, \bar{S}(\not\! p)\, P_R h^T|\ .
\end{equation}
Then,  based on  the  unitarity relation,  also  known as the  optical
theorem, governing the resummed propagators, {\em i.e.}
\begin{equation}
  \label{OT}
S^{-1}(\not\! p)\ -\ S^{-1\dagger}(\not\! p) \ =\ -i \int
d{\rm LIPS} \not\! p (h^T h^* P_L\,  +\,  h^\dagger h P_R)\ ,
\end{equation}
we can prove the equality
\begin{equation}
  \label{Uni}
\int d{\rm LIPS}\
|{\cal T}^{\rm res}_{L\Phi^\dagger \to L\Phi^\dagger, L^C\Phi}|^2\ 
=\ \int d{\rm LIPS}\
|\overline{\cal T}^{\rm res}_{L^C\Phi \to  L\Phi^\dagger,
L^C\Phi}|^2\ .
\end{equation}
Indeed, using Eq.\ (\ref{OT}), we find 
\begin{eqnarray}
  \label{Tres1}
\int d{\rm LIPS}\
|{\cal T}^{\rm res}_{L\Phi^\dagger \to L\Phi^\dagger, L^C\Phi}|^2
&& \nonumber\\
&&\hspace{-4cm}= \int d{\rm LIPS}\
hP_R\, S(\not\! p)\,  \not\! p (h^T h^* P_L\,  +\,  h^\dagger h P_R)\,
S^\dagger (\not\! p)\, P_L h^\dagger\nonumber\\
&&\hspace{-4cm}= -i\, hP_R\, [S(\not\! p)\, -\, S^\dagger (\not\! p)]\, 
P_L h^\dagger\ =\ 2\, hP_R\, \Im m S(\not\! p)\, P_Lh^\dagger\, ,
\end{eqnarray}
and for the CP-conjugate total rate, 
\begin{eqnarray}
  \label{Tres2}
\int d{\rm LIPS}\
|{\cal T}^{\rm res}_{L^C\Phi \to L\Phi^\dagger, L^C\Phi}|^2
& =&   2\, h^*P_L\, \Im m \bar{S}(\not\! p)\, P_Rh^T\, \nonumber\\
&=& 2\, hP_R\, \Im m S(\not\! p)\, P_Lh^\dagger\, .
\end{eqnarray}
As the RHSs  of Eqs.\ (\ref{Tres1}) and  (\ref{Tres2}) are equal,  the
equality  (\ref{Uni}) is then  obvious.   Subtracting Eq.\ (\ref{CPT})
from  Eq.\ (\ref{Uni}), it is not  difficult to  see that $\Delta_{\rm
  CP}$ vanishes at the one-loop resummed level.  We should remark that
unlike  \cite{BP},   the   resummation approach  under   consideration
\cite{AP} satisfies  CPT and unitarity  {\em exactly} without recourse
to  any re-expansion of the resummed  propagator.   If we also include
resummed amplitudes subleading in  the Yukawa couplings, then residual
CP-violating terms which are  formally of order $h^8_{li}$  and higher
occur  in $\Delta_{\rm CP}$.  These terms result from the interference
of two resummed amplitudes containing one-loop vertex graphs.  Because
of unitarity   however,  the residual    CP-violating terms  of  order
$h^8_{li}$ and  $h^{10}_{li}$ will cancel  at  two loops together with
respective     CP-violating  terms  coming   from   one-loop $2\to  4$
scatterings, and so on.

We close   this section by  making  a few important comments.   In the
approach considered  here \cite{AP}, the physical transition amplitude
is   obtained by sandwiching  the  resummed propagators between matrix
elements related to initial and final  states of the resonant process. 
Therefore, diagonalization  of $S (\not\!  p)$  is no  more necessary,
thereby      avoiding    possible    singularities   emanating    from
non-diagonalizable     (Jordan-like)     effective  Hamiltonians   [or
equivalently     $S^{-1} (\not\!\!  p)$]   \cite{AP}.    In fact, such
effective Hamiltonians represent situations  in which the CP-violating
mixing between  the  two unstable  particles  reaches its maximum  and
physical CP  asymmetries  are therefore large.  In   such a case,  the
complex mass  eigenvalues  of the   effective Hamiltonian are  exactly
equal. How    can   then this  observation  be   reconciled  with  the
CP-invariance  condition (\ref{CPinv})?  To   answer this question, we
should notice  that  in the presence  of a  large particle mixing, the
mass eigenstates  of $S^{-1} (\not\!\!   p)$ are generally non-unitary
among  themselves, whereas   the   OS-renormalized  mass   eigenstates
\cite{KP}  form  a    well-defined  unitary   basis    (or  any  other
renormalization scheme  that  preserves orthonormality of the  Hilbert
space),  upon  which perturbation theory can   be  formulated order by
order.   Therefore, the  field-theoretic  OS renormalized  masses  are
those  that  enter the  condition   of CP   invariance  given in  Eq.\ 
(\ref{CPinv}).  Consequently,  if the two  complex mass eigenvalues of
the effective Hamiltonian are  equal, this does not necessarily entail
an equality   between  their respective OS   renormalized masses, and,
hence, absence of CP violation as well \cite{AP}.

\section{Low-energy constraints}

It is  interesting  to look briefly  at  the possible phenomenological
implications of the  leptogenesis model for collider \cite{PRODpp} and
lower energies.  However,   if the out-of-equilibrium constraints  are
imposed on   {\em  all} lepton  families  of  the model,  all possible
new-physics phenomena are estimated to be  extremely small at the tree
or    one-loop level.  Since     sphaleron interactions   preserve the
individual quantum numbers $B/3-L_i$,  {\em e.g.}, $B/3  - L_e$, it is
sufficient to produce the  observed asymmetry in  $B$ by an  excess in
$L_e$.  Most interestingly, the  so-generated  BAU will not be  washed
out,  even  if operators   that violate $L_\mu$  and $L_\tau$   are in
thermal   equilibrium   provided  that    possible   $L_e-L_\mu$-  and
$L_e-L_\tau$-violating interactions  are  absent  \cite{Dreiner/Ross}.
This  is  not very hard  to  imagine. For example,   one  may have two
right-handed   neutrinos that mix with   the electron  family only and
produce the BAU,  and the remaining  isosinglet neutrinos strongly mix
with the $\mu$  and  $\tau$ families.  This   class of heavy  Majorana
neutrino  models may predict   sizeable  new-physics phenomena  due to
significant non-decoupling quantum effects on lepton-flavour-violating
decays of the  $Z$ boson \cite{KPS_Z}, the   $\tau$ and $\mu$  leptons
\cite{IP}.   They can also  give rise to  breaking  of universality in
leptonic  diagonal $Z$-boson \cite{BKPS}   and $\pi$ decays \cite{KP},
{\em etc.} \cite{LL}.

Below the critical temperature  $T_c$  of the first-order  electroweak
phase transition,   the Higgs  boson acquires  a  non-vanishing vacuum
expectation value  $v$ which leads  to non-zero light neutrino masses. 
However, for the  models under discussion, light-neutrino masses scale
as $m_\nu \sim  (v^2\mu_L)/M^2$,   and therefore satisfy easily    the
cosmological upper bound of 10 eV, even if $M$ is as low as 1 TeV.
\begin{center}
\begin{picture}(200,120)(0,0)
  \SetWidth{0.8}

\ArrowLine(50,80)(80,80)\Text(50,90)[l]{$e^-$}
\Line(80,80)(110,80)\Text(95,80)[]{{\boldmath $\times$}}
\Text(95,68)[]{$N_i$}
\ArrowLine(140,80)(110,80)\Text(125,90)[]{$e^-$} 
\Line(140,80)(170,80)\Text(155,80)[]{{\boldmath $\times$}}
\Text(155,92)[]{$N_j$}
\ArrowLine(170,80)(200,80)\Text(200,90)[r]{$e^-$}
\DashArrowArcn(110,80)(30,180,360){5}
\DashArrowArc(140,80)(30,180,270){5}
\DashArrowArc(140,80)(30,270,360){5}
\Photon(140,50)(140,20){3}{4}\Text(145,20)[l]{$\gamma$}
\Text(110,115)[b]{$\chi^-$}
\Text(118,60)[tr]{$\chi^-$}\Text(163,60)[lt]{$\chi^-$}
\end{picture}\\[-0.2cm]
{\small {\bf Fig.\ 3:} ~Typical two-loop diagram contributing to the EDM 
of electron.}
\end{center}

The leptogenesis model  gives rise to an EDM  of electron at two loops
\cite{Ng2}, on which the tight experimental bound $(d_e/e) < 10^{-26}$
cm  \cite{PDG} applies.  A typical  diagram  is shown in   Fig.\ 3.  A
first estimate for $m_{N_2},\ m_{N_1}\gg M_W$ yields
\begin{equation}
\label{EDMaj}
(\frac{d_e}{e}\Big) \sim
(10^{-24}\, \mbox{cm})\, \Im m(h_{1e}h^*_{2e})^2\, 
\frac{m_{N_1}m_{N_2}(m^2_{N_1}-m^2_{N_2})}{(m^2_{N_1} + m^2_{N_2})^2}\, 
\ln\Big(\frac{m_{N_1}}{M_W}\Big)\, ,
\end{equation}
The  above EDM contribution is  several  orders of magnitude below the
experimental bound,  if  $x_N =  (m_{N_2}/m_{N_1}) -  1  < 10^{-3}$ or
$|h_{l1}|,\ |h_{l2}| \stackrel{\displaystyle <}{\sim} 10^{-2}$.

\section{Finite temperature effects and Boltzmann equations}

In the leptogenesis scenario based  on mixing of heavy neutrinos,  one
has to worry about other effects, which might,  to some extend, affect
the resonant condition of CP  violation in Eq.\ (\ref{CPcond}).  Apart
from  the intrinsic  width  of   a  particle resonance,  there may  be
broadening  effects  at  high temperatures  due   to collisions  among
particles.  Such effects will contribute terms  of order $h^4_{li}$ to
the $N_i$ widths and are small in general \cite{Roulet}.  On the other
hand, finite temperature effects on  the $T=0$ masses of particles may
be significant.  Because of the SM gauge interactions, the leptons and
the  Higgs   fields    receive   appreciable    thermal   masses    to
\cite{HAW}, {\em i.e.}
\begin{eqnarray}
\label{thermal}
\frac{m^2_L(T)}{T^2} &=& \frac{1}{32}\, (3g^2\, +\, g'^2)\ \approx\
0.044\, .
\end{eqnarray}
where $g$ and $g'$ are  the SU(2)$_L$ and  U(1)$_Y$ gauge couplings at
the running scale $M_Z$.  The isosinglet  heavy neutrinos also acquire
thermal masses through Yukawa interactions \cite{HAW}, {\em i.e.}
\begin{equation}
\label{mN(T)}
\frac{m^2_{N_i}(T)\, -\, m^2_{N_i}(0)}{T^2}\ =\ \frac{1}{16}\, |h_{li}|^2\, .
\end{equation}
Such a $T$-dependent  mass shift is small  and comparable to the $N_i$
widths at $T\approx  m_{N_i}$.  Therefore, it is easy  to see that the
condition    for   resonant  CP violation     through   mixing in Eq.\ 
(\ref{CPcond}) is qualitatively  satisfied.  Finally, the Higgs  field
also receives   appreciable  thermal contributions.   The  authors  in
\cite{CKO} have calculated the  one-loop Higgs thermal mass. They find
that  $M_\Phi   (T)/T \stackrel{\displaystyle  <}{\sim} 0.6$  for LEP2
favoured values of the Higgs-boson mass, {\em i.e.}, $M_H  < 200$.  In
this  Higgs-mass range,  the effective    decay  widths of the   heavy
neutrinos   $\Gamma_{N_i}(T)$   will     be   reduced    relative   to
$\Gamma_{N_i}(0)$  by a  factor   of two or    three due to   sizeable
phase-space corrections.  Nevertheless, the  influence on the resonant
phenomenon  of CP  violation through mixing  is not  dramatic when the
latter  effects  are included,    and  therefore large   leptonic   CP
asymmetries can still be generated.

Before we   present  plots of  numerical estimates,  it  is  useful to
discuss  first the out-of equilibrium   constraints on heavy  neutrino
decays \cite{KT}. These constraints are qualitatively given by
\begin{equation}
\label{Sakh3}
\Gamma_{N_i} (T=m_{N_i})\ \stackrel{\displaystyle <}{\sim}\
2K\, H(T=m_{N_i})\, ,
\end{equation}
where $K\approx 1 - 1000$ is a factor that quantifies how far from the
expansion rate of Universe  heavy-neutrino lifetimes are and $H(T)$ is
the Hubble parameter
\begin{equation}
\label{Hubble}
H(T)\ =\ 1.73\, g_*^{1/2}\, \frac{T^2}{M_{Planck}}\ ,
\end{equation}
with $M_{Planck}  = 1.2\ 10^{19}$ GeV   and $g_*\approx 100-400$ being
the number of active degrees of freedom in usual extensions of the SM.
Then, the  out-of    equilibrium constraint  in   Eq.\   (\ref{Sakh3})
translates into the bound
\begin{equation}
\label{hli_bound}
|h_{li}|^2\ \stackrel{\displaystyle <}{\sim}\
7.2 K \times 10^{-14}\, \Big( \frac{m_{N_i}}{1\ \mbox{TeV}}\Big)\, .
\end{equation}
The above bound is considered for all Yukawa couplings. 

Above  the   electroweak    phase  transition,  the    $B+L$-sphaleron
interactions convert  approximately one-third of the lepton-to-entropy
density ratio $Y_L  =  n_L/s$ into a baryon-to-entropy   density ratio
$Y_B = n_B/s$, {\em i.e.}  \cite{BS,HT}
\begin{equation}
\label{YB_YL}
Y_B\approx -\, \frac{1}{3}\, Y_L\, \approx -\,\frac{1}{3K}\
\frac{\delta_{N_i}}{g_*}\ .
\end{equation}
The last  approximate equality represents  the asymptotic  solution of
the relevant Boltzmann  equations (BEs) \cite{MAL,APRD}.   {}From Eq.\ 
(\ref{YB_YL}), we  see that $Y_B$ can  be in observed ball  park, {\em
  i.e.}\ $Y_B\approx 10^{-10}$,  if  $|\delta_{N_i}|/K$  are of  order
$10^{-7} - 10^{-6}$. Clearly, CP asymmetries of  order unity allow for
very large values  of $K$.  As  a consequence, the thermal plasma  can
then be rather dense and the  conditions of kinetic equilibrium in BEs
can  comfortably be satisfied   even  within the minimal  leptogenesis
scenario.
\begin{figure}[ht]
   \leavevmode
 \begin{center}
   \epsfxsize=10.cm
   \epsffile[0 0 539 652]{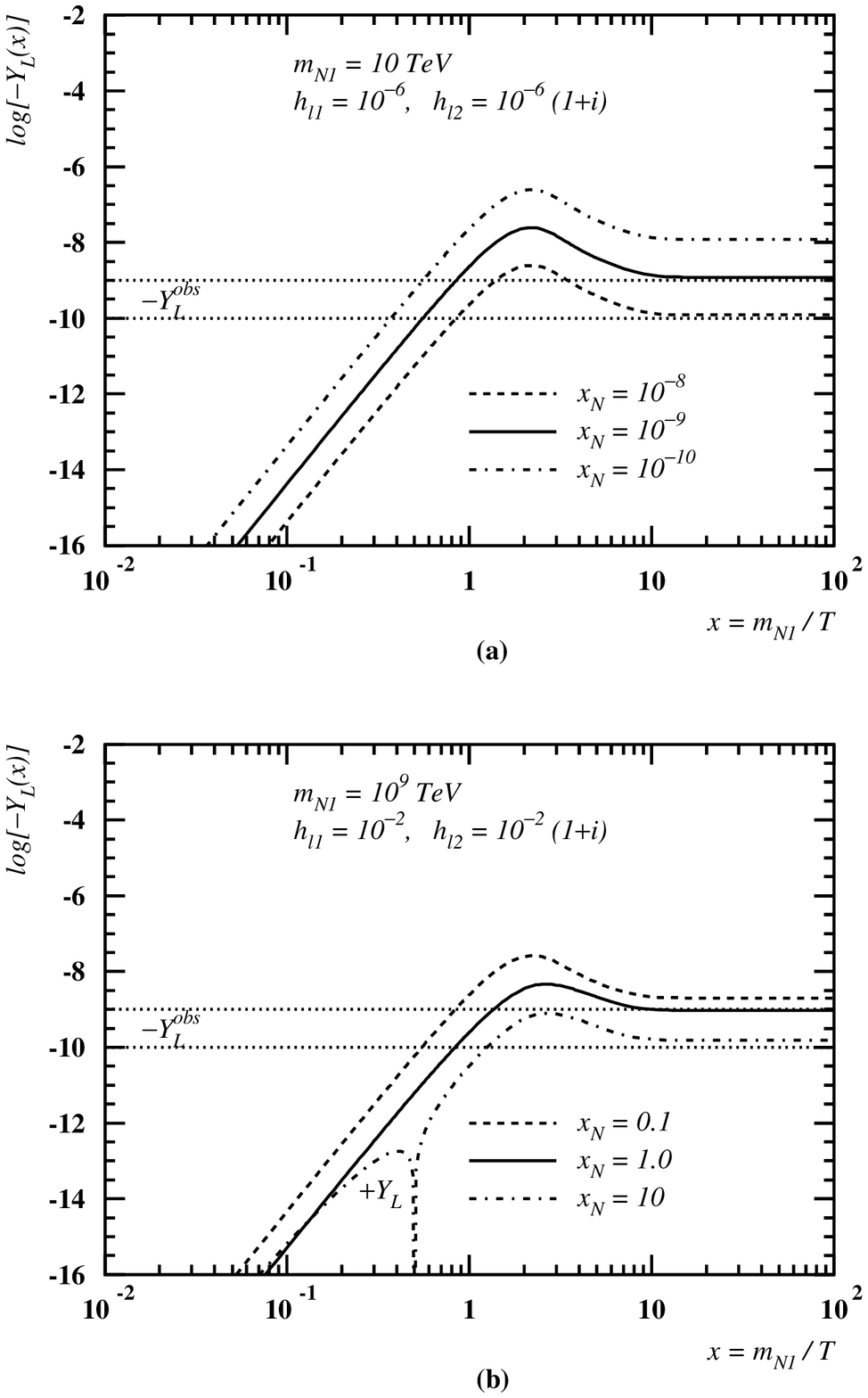}
{\small {\bf Fig.\ 4:} Lepton asymmetries for selected heavy Majorana 
neutrino scenarios.}
 \end{center}
\end{figure}

In  Fig.\ 4, we plot  the dependence of  $Y_L(x)$ on $x=m_{N_1}/T$ for
two representative    scenarios,  after solving  numerically   the BEs
\cite{APRD}.  The observed range  for $Y_L$, $Y^{obs}_L$, is indicated
with two  confining horizontal  dotted lines.   The parameter  $x_N  =
m_{N_2}/m_{N_1} - 1$ is a measure of the degree of mass degeneracy for
$N_1$  and $N_2$.  In  the first scenario in  Fig.\ 4(a), $m_{N_2} \ge
m_{N_1} = 10$ TeV, and  $h_{l1} = 10^{-6}$,  $h_{l2} = 10^{-6} (1+i)$. 
In this case, a heavy-neutrino mass  splitting $x_N$ of order $10^{-9}
- 10^{-8}$ is  sufficient to account for the  BAU.  We  also find that
the  $\varepsilon$-type CP-violating  mechanism  is  dominant, whereas
$\varepsilon'$-type CP violation is  extremely suppressed.  The second
scenario in Fig.\ 4(b)  is close to the  traditional one considered in
\cite{FY}, where  $m_{N_1}  = 10^9$ TeV, and  $h_{l1}   = 10^{-2}$ and
$h_{l2} = 10^{-2}  (1 + i)$.  Obviously, it  is not mandatory  here to
have a high degree of degeneracy  for $N_1$ and  $N_2$ in order to get
sufficient CP violation for the BAU. In this case, both $\varepsilon$-
and $\varepsilon'$-   type  mechanisms of  CP  violation  are  equally
important.

\section{Conclusions}

We have considered  the scenario of baryogenesis through  leptogenesis
in which  out-of-equilibrium  $L$-violating decays  of heavy  Majorana
neutrinos produce an   excess in $L$,  which   is converted into   the
observed asymmetry   in  $B$, through  the  $B+L$-violating  sphaleron
interactions.   In  particular,  we have studied   the impact  of  the
$\varepsilon$- and $\varepsilon'$-type  mechanisms for CP violation on
generating  the required excess of baryons  detected in  the Universe. 
Based on the effective resummation approach outlined  in Section 3, we
have found that $\varepsilon$-type CP violation is resonantly enhanced
if the mass splitting of the heavy Majorana neutrinos is comparable to
their widths  ({\em c.f.}\ Eq.\  (\ref{CPcond})), and if the parameter
$\delta_{CP}$ defined  in Eq.\ (\ref{dCP}) has  a value  close to one. 
These are the two necessary and sufficient  conditions for resonant CP
violation of order unity \cite{APRD}.   As a consequence, the scale of
lepogenesis   may  be  lowered  up   to  the  TeV   range.   In  fact,
E$_6$-motivated scenarios   with  nearly  degenerate    heavy Majorana
neutrinos and all Yukawa  couplings being of  the same order can still
be  responsible   for   the BAU.    In     this kinematic   range, the
$\varepsilon'$-type    contributions  are  extremely suppressed.  Even
though interesting alternatives  have been proposed  in the literature
\cite{MS}, the  leptogenesis    model considered  here   is  the  most
economical  one.  Just   two   isosinglet neutrinos  are    needed for
baryogenesis. Finally, we  must remark that finite temparature effects
and constraints  due to electron EDM  have a rather marginal effect on
CP asymmetries.

\end{document}